\documentstyle[12pt]{article}
\def\roughly#1{\,\,\raise.3ex\hbox{$#1$\kern-.75em\lower1ex\hbox{$\sim$}}\,\,}
\def\lappeq{\roughly{<}}
\def\gappeq{\roughly{>}}

\begin{document}
\begin{titlepage}
\begin{center}
October 1996\hfill    UND-HEP-96-US02 \\
               \hfill    hep-ph/9610341
\vskip .2in
{\large \bf
Precision Top Mass Measurements vs. Yukawa Unification Predictions\footnote{to appear in Proceedings of the 1996 DPF/DPB Summer Study on New Directions for High-Energy Physics (Snowmass 96), June 24 - July 12 1996, Snowmass, Colorado.}}
\vskip .3in
Uri Sarid\footnote{E-mail:
sarid@particle.phys.nd.edu}\\[.03in]
{\em Department of Physics\\
     University of Notre Dame\\
     Notre Dame, IN 46556}
\end{center}
\vskip .2in
\begin{abstract}
\medskip
How accurately should the top quark mass be measured in order to test theoretical predictions? A possible answer is presented within a particular theoretical framework, that of top-bottom-tau Yukawa unification in a supersymmetric SO(10) grand unified theory. Yukawa unification, and the uncertainties in its $m_t$ prediction, are introduced by analogy to gauge unification and the uncertainties in the predictions of $\sin^2\theta_W$ or $\alpha_3(m_Z)$. There are two sources of uncertainty in this framework: ``removable'' uncertainties due to physics at the electroweak and supersymmetry-breaking scales, and ``irremovable'' ones from physics at and above the unification scale. The latter are precisely the model-dependent effects which would shed light on the nature of the unified model, so they may be regarded as a (model-dependent) part of the prediction rather than as uncertainties. The removable sources are estimated first using current experimental bounds, and then using plausible guesses for the bounds that may be available within roughly a decade: they are not likely to be reduced below roughly $\pm 1\,\rm GeV$. That is the level at which such unified theories will be testable against future experimental determinations of the top mass. 
\end{abstract}
\end{titlepage}

\section{Introduction}
At this Snowmass meeting, experimental proposals have been discussed which would attempt to measure the mass of the top quark very precisely, to perhaps a few hundred MeV, in a timeframe of order a decade. It is of interest then to investigate how precisely can $m_t$ be predicted theoretically, and thereby estimate what we would learn about various theories by the comparison of predictions and experimental results. In this brief report I discuss one particular context in which $m_t$ can be predicted. Most of the results I will present are based upon the detailed investigations carried out with my collaborators L.J.~Hall and R.~Rattazzi in Refs.~\cite{ref:toppred,ref:gbu,ref:rewsm} (see also Ref.~\cite{ref:altpred}), to which the reader is referred for further details. While the context I will focus on will be complete (top-bottom-tau) Yukawa unification, the qualitative conclusions and to some extent also the quantitative lower bounds on the precision of the predictions are much more generic, at least in unified scenarios: the central value of the predictions often (fortunately!) depend on the models, but many of the uncertainties are expected to be more model-independent.

Roughly speaking, measurements of the top mass can test Yukawa unification \cite{ref:yukuni} in the same sense that measuring the weak mixing angle $\sin^2\theta_W$, or better yet the strong coupling $\alpha_3$, can test gauge unification. Let us first recall the case of gauge unification, in order to illustrate how the unification hypothesis is tested. In unified gauge theories, the standard-model gauge group $G_{\rm SM} = \rm SU(3)\times SU(2)\times U(1)_Y$ is usually embedded in a simple group $G_U$ such as SU(5) or SO(10); the former is the simplest, smallest group that can accommodate $G_{\rm SM}$, while SO(10) [which contains SU(5) as a subgroup] is somewhat bigger but can unify an entire family of quarks and leptons in a single irreducible representation. The tree-level prediction of such theories is the equality of the three gauge couplings: $g_1 = g_2 = g_3$, where $g_1^2 = {5\over3} g_Y^2$ is properly normalized to correspond to a generator of the unified gauge group. The weak mixing angle is then predicted to be $\sin^2\theta_W = 3/8$. This seemingly incorrect prediction is significantly changed by radiative corrections if the unified group $G_U$ is spontaneously broken at a large scale $M_U \gg m_Z$. The leading $\log (M_U/m_Z)$ effects are then summed using the renormalization group (RG) evolution of the gauge couplings, which introduces a dependence on the particle content of the theory between $M_U$ and $m_Z$. (I will not discuss theories with intermediate scales in this work.) It is at this stage that the $\sin^2\theta_W \simeq 0.23$ prediction of the minimal supersymmetric extension of the standard model, the MSSM, is strongly favored over the $\sin^2\theta_W \simeq 0.215$ prediction of the standard model alone. To obtain further precision, 2-loop RG equations may be used, in which case 1-loop threshold corrections --- which are of the same order --- must also be added, and they introduce a dependence on the masses and couplings of the theories at the high scale $\sim M_U$ and at the low scale $\sim m_Z$. 

To this precision, the comparison between theory and experiment (both given in the $\overline{\rm MS}$ scheme at the scale $m_Z$) reads:
\begin{eqnarray}
\sin^2\theta_W^{\rm pred} &=& 0.2357 \pm 0.0014 \pm 0.004 
\nonumber\\
\sin^2\theta_W^{\rm expt} &=& 0.2319 \pm 0.0005 \,.
\label{eq:ssth}
\end{eqnarray}
(These values are not necessarily the latest and most authoritative ones available, but are sufficient for our purposes.) The first uncertainty in the prediction is mainly due to experimental uncertainty in the input value of $\alpha_3(m_Z)$ and to uncertainty in the masses of the superpartners, though all are assumed to lie roughly below a TeV. The second theoretical uncertainty is harder to estimate, originating in threshold corrections at the GUT scale $M_U$ and the presence of higher-dimensional operators suppressed by inverse powers of a higher scale such as the Planck mass. Since the experimental value of $\sin^2\theta_W$ is so well measured, it is more convenient to reverse the calculation which led to Eq.~(\ref{eq:ssth}) and use $\sin^2\theta_W^{\rm expt}$ to make a GUT prediction of $\alpha_3(m_Z)$:
\begin{eqnarray}
\alpha_3^{\rm pred} &=& 0.132 \overbrace{\,\pm\,\,0.004\,}^{\rm removable\,\,} \overbrace{\phantom{.}\pm\, 0.015\,}^{\rm irremovable}
\nonumber\\
\alpha_3^{\rm expt} &=& 0.117 \pm 0.005 \,.
\label{eq:alphs}
\end{eqnarray}
(See the disclaimer above.) For our purposes the uncertainties are more relevant than the central values. The hope and the expectation is that, within a decade, $\alpha_3^{\rm expt}$ and at least some of the superpartner masses will be known sufficiently well that all the uncertainties associated with physics at scales $\sim m_Z$ would be insignificant relative to those stemming from scales $\gappeq M_U$. I will call the former ``removable'' uncertainties and the latter ``irremovable'' since only the former are likely to be directly confronted and reduced by experiments. Once the removable uncertainties are reduced, then comparing the prediction of $\alpha_3$ with the experimental value will be a measurement, within the GUT context, of the irremovable effects, which is really what is meant by a measurement of gauge unification: the unification of gauge couplings at a very high scale, within a given theory characterized by certain masses and couplings and higher-dimensional operators at that scale.

\section{Yukawa Unification}
In SO(10) models, each entire generation of quark and lepton superfields (including a right-handed neutrino) is perfectly contained in a single, 16-dimensional irreducible representation of $G_U$, and the two Higgs doublets needed in supersymmetric models to give up-and down-type quarks masses can fit, along with a pair of triplets, in a single 10-dimensional irrep. While the light two generations of quarks and leptons require more structure to explain their masses and mixings, the third generation masses may be well described by a single, large Yukawa coupling $\lambda_G \underbar{16}_3 \underbar{10}_H \underbar{16}_3$. [By ``Yukawa unification'' I will mean this full top-bottom-tau unification, rather than just the bottom-tau unification \cite{ref:btau} of earlier SU(5) models.] The picture that results from this assumption of (third-generation) Yukawa unification is appealing in its simplicity and resemblance to gauge unification: while $\lambda_t = \lambda_b = \lambda_\tau = \lambda_G$ at the scale $M_U$, below this scale the three Yukawa couplings evolve differently in the non-SO(10)-symmetric effective theory. The low-energy Yukawa couplings yield quark masses when the Higgs doublets $H_U$ and $H_D$ acquire vacuum expectation values (VEVs) $v_{U,D}$, where $v_U^2 + v_D^2 \simeq (174\,\rm GeV)^2$. In gauge unification, one measures two quantities at low energies, say $\sin^2\theta_W$ and $\alpha_{\rm em}$, to fix the high-energy parameters $M_U$ and $\alpha_U$; the remaining low-energy quantity $\alpha_3$ is then predicted. In Yukawa unification, there are four low-energy quantities, namely the three Yukawa couplings and the ratio $\tan\beta = v_U/v_D$, related to two high-energy parameters, namely $\lambda_G$ and a Higgs-sector parameter (or combination of parameters) which determines the form of electroweak symmetry breaking; the unification scale is already fixed by gauge unification. Thus $m_b$ and $m_\tau$ are input from experiment, and $m_t$ and $\tan\beta$ are predicted. 

The top-bottom mass hierarchy in such models results from a Higgs-VEV hierarchy: since $\lambda_t$ and $\lambda_b$ remain comparable at all scales, $m_t/m_b = (\lambda_t v_U)/(\lambda_b v_D) \sim \tan\beta \sim 50$. Generation of such a large hierarchy, at least in the usual models of hidden-sector supersymmetry breaking communicated to the MSSM at Planckian scales, favors \cite{ref:rewsm} a rather specific hierarchical superpartner spectrum needed to make the theory most natural (though a fine-tuning of order $\sim 1/\tan\beta$ remains \cite{ref:rewsm,ref:nr}). If the spectrum is not strongly hierarchical, threshold corrections $\delta m_b/m_b$ to the bottom quark mass introduce a strong power-law dependence of $m_t^{\rm pred}$ on the spectrum \cite{ref:toppred} (see also Ref.~\cite{ref:hempone}), so in any case the superspectrum is more intimately involved in Yukawa unification than in gauge unification. Therefore experimental determination of this spectrum will be crucial in testing the Yukawa unification hypothesis. (There are other tests of this hypothesis, such as predictions of $b\to s\gamma$ \cite{ref:toppred,ref:gbu,ref:rewsm,ref:altbsg}, but we will focus on the top mass in this report.) 

\section{Uncertainties}
I will assume in the following that within the next decade or so the superspectrum will be roughly mapped out, either through the discovery of most of the superpartners or the determination that the squarks are much heavier than the higgsinos and charginos. If instead all superpartner masses $\widetilde m_i$ remain beyond experimental reach, then the relevance of supersymmetry itself is questioned, while the above-mentioned $\delta m_b/m_b$ corrections cannot be directly measured. I will also assume that we will be able to translate with sufficient precision the $\overline{\rm MS}$ top mass prediction, or the related pole mass prediction $m_t^{\rm pole}$, into the experimentally-measured top mass. Making these assumptions, then, cleanly separates the uncertainties in the prediction of $m_t$ into {\bf removable} sources:
\begin{enumerate}
\item the bottom quark mass $m_b$;
\item the strong coupling $\alpha_3$;
\item the potentially large, finite threshold corrections $\delta m_b/m_b$;
\item other threshold corrections, usually $\sim \log(\widetilde m_i/m_Z)$;
\end{enumerate}
and {\bf irremovable} sources:
\begin{enumerate}
\item high-energy thresholds $\sim \log(M_i/M_U)$ where $M_i$ are GUT-scale masses;
\item higher-dimensional operators (as in gauge unification);
\item $\lambda_t(M_U) \ne \lambda_{b,\tau}(M_U)$ in certain models.
\end{enumerate}
The last possibility exists, for example, in some models having large mixings between the second and third generations, or when the MSSM Higgs doublets contain significant admixtures of more than a single $\underbar{10}_H$ representation.

The details of the central values --- typically 170--180 GeV --- and the uncertainties in the top mass prediction can be found in Ref.~\cite{ref:toppred}; as before, it is the uncertainties that are of interest here. In the table I have summarized the current uncertainties in the input parameters $\Delta_{\rm now}$ and the resulting uncertainties in $m_t^{\rm pred}$. The irremovable uncertainties (denoted by ``GUT Thr.'' in the table), which include the above three sources, are only estimates. In any case they are the very effects one is trying to measure, because they are just as much a prediction of any particular Yukawa-unified models as the ``central'' value, and are only listed as uncertainties because they are more model-dependent. They will be eliminated (that is, the central value will be shifted and fixed) only when some particular model is chosen. But in a decade or so many of the removable uncertainties will be reduced, at least if the superspectrum is partially characterized. It is also hoped \cite{ref:futalphs} that the uncertainty in $\alpha_3$ will be improved by almost an order of magnitude. And perhaps a combination of lattice results, QCD sum rules and a better understanding of other QCD-related issues such as renormalons would result in a much improved (and very important!) determination of $m_b(m_b)$, the running bottom quark mass defined at its own mass scale. (I choose an uncertainty of 15 MeV because that's roughly half of the most optimistic uncertainties quoted today, and an order of magnitude smaller than the most conservative ones; thus it is perhaps a fair reflection of what at least some theorists believe is possible to achieve with existing theoretical methods.) All these plausible guesses are shown as $\Delta_{\rm fut.}$ in the table. The ranges of $m_t$ uncertainties (such as $\pm 0.5 - 2 \,\rm GeV$) arise because the uncertainties depend on the central values of the various parameters, that is, $m_t$ may be more or less sensitive to a given parameter when that parameter (or others) is large or small. 

\begin{table}[h]
\begin{center}
\caption{Current and future uncertainties in the $m_t$ prediction.}
\medskip
\label{tab:unc}
\begin{tabular}{|r|ll|ll|}
\hline
\hline
Input & $\!\Delta_{\rm now}\Rightarrow$ & 
$\!\!\!\!\Delta m_t/\rm GeV$
& $\!\Delta_{\rm fut.}\Rightarrow$ & 
$\!\!\!\!\Delta m_t/\rm GeV$\\
\hline
$m_b/\rm MeV$ & $\pm 200$ & $\pm 7$ 
  & $\pm 15$ & $\pm 0.5$\\
$\alpha_3(m_Z)$ & $\pm 0.005$ & $\pm 3-10$ 
  & $\pm 0.001$ & $\pm 0.5-2$\\
$\delta m_b/m_b$ & small? & ? 
  & $\pm 10\%$ & $\pm 0.5-2$\\
$\log\,\tilde m_i$ & $\pm (\lappeq 3)$ & $\pm 5-10$ 
  & $\pm 10\%$ & $\pm (\lappeq 0.5)$\\
GUT thr.& ? & $\pm$ few? 
  & ? & $\pm$ few?\\
\hline
\hline
\end{tabular}
\end{center}
\end{table}

\section{Conclusions}
We learn from this table that, even in favorable circumstances, the ``removable'' uncertainties due to low-energy measurable parameters cannot be reduced much below the $\sim 1\,\rm GeV$ level. Thus a more precise measurement of $m_t$, say to within a few hundred MeV, does not appear necessary in order to measure the high-energy unification of all three Yukawa couplings. An experimental measurement of $m_t$ to within $\sim 1\,\rm GeV$ in the next decade would be sufficient, and by the time such a precision is reached, the removable uncertainties may well be reduced to the same level, making $m_t$ a very useful probe of the degree of Yukawa unification at the scale $M_U$, and allowing discrimination (albeit indirect) between various Yukawa-unified models. 

One final note: within a few years we may well be able to determine whether $\tan\beta$ is large or small from measurements of the chargino and neutralino properties \cite{ref:expltb}, even if a precise value of $\tan\beta$ is not yet available. It would be sufficient to know that $\tan\beta > 13$ to conclude, using $\sin\beta > 0.997$, that $\lambda_t$ is determined by $m_t$ to within a third of a percent. Then a precise measurement of $m_t$ would amount to an almost direct measurement of the Yukawa coupling $\lambda_t$ itself. Such a measurement would have interesting consequences for a wide range of models, not just those unifying the third-generation Yukawa couplings.

%%%%%

%%%%% References
%


\begin{thebibliography}{2}
  
\bibitem{ref:toppred} L.J.~Hall, R.~Rattazzi and U.~Sarid, Phys.\ Rev.\ D {\bf 50}, 7048 (1994).

\bibitem{ref:gbu} R.~Rattazzi, U.~Sarid and L.J.~Hall, SU-ITP-94-15, RU-94-37, {\it Proceedings of the Second IFT Workshop on Yukawa Couplings and the Origins of Mass} (1994).

\bibitem{ref:rewsm} R.~Rattazzi and U.~Sarid, Phys.\ Rev.\ D {\bf 53}, 1553 (1996).

\bibitem{ref:altpred} M. Carena, M. Olechowski, S. Pokorski and
C.E.M. Wagner, Nucl.\ Phys.\ {\bf B426}, 269 (1994); M. Carena and C.E.M. Wagner, CERN-TH-7321-94, {\it Proceedings of the Second IFT Workshop on Yukawa Couplings and the Origins of Mass} (1994)

\bibitem{ref:yukuni} G.F. Giudice and G. Ridolfi, Z.\ Phys.\ C {\bf 41}, 447 (1988); M. Olechowski and S. Pokorski, Phys.\ Lett.\ B {\bf214}, 393 (1988); P.H. Chankowski, Phys.\ Rev.\ D {\bf 41}, 2877 (1990); B. Ananthanarayan, G. Lazarides, and Q. Shafi, Phys.\ Rev.\ D {\bf 44}, 1613 (1991); M. Drees and M.M. Nojiri, Nucl.\ Phys.\ {\bf B369}, 54 (1992);  B. Ananthanarayan, G. Lazarides, and Q. Shafi, Phys.\ Lett.\ B {\bf 300}, 245 (1993); H. Arason, D.J. Casta\~{n}o, B.E. Keszthelyi, S. Mikaelian, E.J. Piard, P. Ramond, and B.D. Wright, Phys.\ Rev.\ Lett.\ {\bf 67}, 2933 (1991); S. Kelley, J.L. Lopez, and D.V. Nanopoulos, Phys.\ Lett.\ B {\bf 274}, 387 (1992); V. Barger, M.S. Berger, and P. Ohmann, Phys.\ Rev.\ D {\bf 47}, 1093 (1993).

\bibitem{ref:btau} M. Chanowitz, J. Ellis, and M.K. Gaillard, Nucl.\ Phys.\ {\bf B135}, 66 (1978).

\bibitem{ref:nr} A.E. Nelson and L. Randall, Phys.\ Lett.\ B {\bf 316}, 516 (1993); R. Hempfling, Phys.\ Rev.\ D {\bf 52}, 4106 (1995).

\bibitem{ref:hempone} R. Hempfling, Phys.\ Rev.\ D {\bf 49}, 6168 (1994).

\bibitem{ref:altbsg} R.~Garisto and J.N.~Ng, Phys.~Lett.~B {\bf 315}, 372 (1993); M.A.~D\'\i az, Phys.~Lett.~B {\bf 322}, 207 (1994); F.M.~Borzumati, Z.\ Phys.\ C {\bf 63}, 291 (1994).

\bibitem{ref:futalphs} S. Kuhlman {\it et al.} (The NLC Accelerator Design Group and the NLC Physics Working Group), preprint BNL 52-502, Fermilab-PUB-96/112, LBNL-PUB-5425, SLAC Report 485, UCRL-ID-124160, UC-414, submitted to this Snowmass '96 Workshop.

\bibitem{ref:expltb} J. Feng, private communication.

\end{thebibliography}
\end{document}